\documentclass[english,structabstract]{aa}
\pdfoutput=1
\usepackage{txfonts}
\usepackage{natbib}
\bibpunct{(}{)}{;}{a}{}{,} % to follow the A&A style
\usepackage{babel}
\usepackage{graphicx}

\begin{document}

\title{Discovery of VVV CL001.}

\subtitle{A Low-Mass Globular Cluster Next to UKS~1 in the Direction of the Galactic Bulge}

\author{D.~Minniti\inst{\ref{inst1}}\and M.~Hempel\inst{\ref{inst1}}\and
  I.~Toledo\inst{\ref{inst1}}\and V.D.~Ivanov\inst{\ref{inst2}}\and  
  J.~Alonso-Garc\'ia\inst{\ref{inst1}}\and R.~Saito\inst{\ref{inst1}}\and
  M.~Catelan\inst{\ref{inst1}}\and D.~Geisler\inst{\ref{inst3}}\and
  A.~Jord\'an\inst{\ref{inst1}}\and J.~Borissova\inst{\ref{inst4}}\and
  M.~Zoccali\inst{\ref{inst1}}\and
R.~Kurtev\inst{\ref{inst4}}\and G.~Carraro\inst{\ref{inst2}}\and 
B.~Barbuy\inst{\ref{inst5}}\and J.~Clari\'a\inst{\ref{inst6}}\and
M.~Rejkuba\inst{\ref{inst7}}\and J.~Emerson\inst{\ref{inst8}}\and
C.~Moni Bidin\inst{\ref{inst3}}}

\institute{Departamento de Astronom\'ia y Astrof\'isica, P. Universidad
  Cat\'olica de Chile, Casilla 306, Santiago, Chile,
  \email{dante,mhempel,itoledoc,jalonso,r.saito,mcatelan,ajordan,mzoccali@astro.puc.cl} \label{inst1}
\and
European Southern Observatory, Alonso de C\'ordova 3107, Vitacura, Casilla 19001, Santiago de Chile 19, Chile, \email{vivanov,gcarraro@eso.org}\label{inst2}
\and
Departamento de Astronom\'ia, Universidad de Concepci\'on, Bio-Bio 160-C,
Chile, \email{dgeisler,cmbidin@astro-udec.cl} \label{inst3}
\and
Departamento de F\'isica y Astronom\'ia, Facultad de Ciencias, Universidad de
Valpara\'iso, Av. Gran Breta\~{n}a 1111, Valpara\'iso, Chile,
\email{jura.borissova,radostin.kurtev@uv.cl}\label{inst4}
\and
Inst Astronomico e Geofisico-Depto Astronomia, Universidade de S\~ao Paulo,
Rua do Mat\~ao 1226, Cidade Universit\'aria, S\~ao Paulo, SP 05508-900,
Brazil, \email{barbuy@astro.iag.usp.br}\label{inst5}
\and
Observatorio Astronomico, Laprida 854, C\'ordoba, 5000, Argentina, \email{jjclaria@gmail.com}\label{inst6}
\and
European Southern Observatory, Karl-Schwarszchild Strasse 2, D85748 Garching/ bei M\H{u}nchen, Germany, \email{mrejkuba@eso.org}\label{inst7}
\and
Astronomy Unit, School of Mathematical Sciences, Queen Mary, University of
London, Mile End Road, London, E1 4NS, UK, \\
\email{j.p.emerson@qmul.ac.uk}\label{inst8}}

\offprints{dante@astro.puc.cl}

\date{Received 20 September 2010; Accepted: 28 November 2010}

\date{}

\abstract{It is not known how many globular clusters may have been left
  undetected towards the Galactic bulge.}  {One of the aims of the VISTA
  Variables in the Via Lactea (VVV) Survey is to accurately measure the
  physical parameters of the known globular clusters in the inner regions of
  the Milky Way and to search for new ones, hidden in regions of large
  extinction.}  {Deep near infrared images give deep $JHK_{\rm S}$-band
  photometry of a region surrounding the known globular cluster UKS~1 and
  reveal a new low-mass globular cluster candidate that we name VVV CL001. }
 {We use the horizontal branch red clump in order to measure
 E(B-V)$\sim$2.2 mag, $(m-M)_0=16.01$ mag, and D=15.9 kpc for the
 globular cluster UKS~1.  Based on the near-infrared colour
 magnitude diagrams, we also measure that VVV CL001 has
 E(B-V)$\sim$2.0, and that it is at least as metal-poor as UKS~1,
 however, its distance remains uncertain.}  {Our finding confirms
 the previous projection that the central region of the Milky Way
 harbors more globular clusters. VVV~CL001 and UKS~1 are good
 candidates for a physical cluster binary, but follow-up
 observations are needed to decide if they are located at the same
 distance and have similar radial velocities. }


\maketitle

\section{Introduction}
\label{Intro}

The inner regions of the Milky Way have been mapped thoroughly at all
wavelengths. Yet, it is not known if there are still some distant globular clusters
awaiting to be discovered, hidden beyond the bulge, due to the high density of stellar sources and the
large and inhomogeneous interstellar extinction. Near-IR surveys have an
advantage for searching these regions. Indeed, the 2MASS discovered two new
globular clusters \citep{Hurt}. But the limiting magnitude of 2MASS
\citep[$K_{\rm S}$=14.3, for 10$\sigma$-detections;][]{Skrutskie06} may prevent the
discovery of fainter objects, especially if they are located in highly
  reddened regions.

The asymmetry of the spatial distribution of known globular
clusters around the Galactic center indicates that previous
observations may have overlooked some additional globular clusters.
\cite{Ivanov2005} recently estimated that there may be about
10 clusters missing towards the inner Milky Way. The recent discoveries (last
10 years) include both faint (low mass) halo clusters as well as reddened
globular clusters projected toward
the bulge, e.g. 2MASS~GC01 and 2MASS~GC02 by \cite{Hurt} 
\citep[see also][]{Ivanov2000}, ESO 280~SC06 by \cite{Ortolani2000}, GLIMPSE~C01 by
Kobulnicky et al. (2005; but see also Ivanov et al. 2005; Davies et al. 2010),
GLIMPSE~C02 by \cite{Kurtev}, AL-3 by \cite{Ortolani2006}, FSR 1735 by \cite{Froebrich}, Koposov~1 and
Koposov~2 by \cite{koposov07}, FSR~1767 by \cite{Bonatto}, Whiting~1
  \citep[][]{carraro05} and Pfleiderer~2 by
\cite{Ortolani2009}.

The VISTA Variables in the Via Lactea (VVV) Public Survey has started mapping
the inner disk and bulge of our Galaxy with VISTA 4m telescope (Visible and Infrared
Survey Telescope for Astronomy) in the
near-IR \citep{Minniti,Saito}. One of the main scientific goals of
the VVV Survey is to study the bulge globular clusters and to search for new
clusters. Here we present VVV~CL001, the first globular cluster candidate discovered by
the VVV Survey.

\section{The VVV Survey Data}

The VVV Survey data are acquired with the VISTA 4m telescope at ESO Paranal
Observatory \citep[][]{Emerson}. The VVV field b351 was observed in the
$JHK_{\rm S}$ bands under subarcsec seeing conditions ($0.8$ arcsec in $K_{\rm S}$). The YZ- band observations are still pending.
Each one of the VVV fields (tiles) covers $1.636\deg^{2}$ in total ($1.475^\circ$
in $l$ by $1.109^\circ$ in $b$). The bulge field b351 observed here is centered
at  $\alpha=17:50:05.42$, $\delta=-23:43:16.7$, $l=4.987^\circ$, $b=1.838^\circ$.

We use here the images processed by CASU VIRCAM
pipeline v1.0 \citep[e.g.][]{irwin04}. The photometry was obtained with DoPhot
\citep{Schechter}. Also, the photometry is uniformly calibrated against the 2MASS
catalog \citep{Skrutskie06}.
The limiting magnitude of the single epoch VVV images is $Ks=18.1$ in the
bulge fields (for details on the observing strategy, see Minniti et
  al. 2010).

The distance probed along the line of sight depends on the
reddening of the fields. For example, in zero reddening disk fields we would
see horizontal branch red clump beyond 50 kpc. Therefore we can search for 
distant galactic globular clusters and measure their physical parameters.

Visual inspection of the images of the field b351 led to the serendipitous discovery of
a star cluster candidate that we name VVV~CL001. This object is located in the vicinity of the known globular 
cluster UKS~1 (Figure \ref{fig1}). Based on near infrared stellar
density maps (see Figure \ref{density})  we conclude that this is not a statistical fluctuation of the background, 
and that the cluster VVV~CL001 is centered at
$\alpha=17:54:42.5$, $\delta=-24:00:53$ 
($l=5.27^\circ$,  $b=0.78^\circ$). 
Applying the procedure by
  \cite[][]{koposov08,belokurov09},  the statistical significance  of the
  over-density at the coordinates given above is S= 9.31. There is no source identified at this
location based on radio, infrared, and visible data on the SIMBAD, 2MASS and NED
databases, or any other star cluster catalog. The second obvious density peak at
  $\alpha=17:54:38$, $\delta=-24:00:49$ is due to a saturated star.

The proximity of VVV~CL001 and UKS~1 on the sky (Figure \ref{fig1}) made it appropriate
to use the latter cluster as comparison in order to measure the reddening and distance of the new
globular cluster candidate. Therefore, we concentrate first on measuring the
parameters of UKS~1 using VVV Survey data.

\section{The Globular Cluster UKS~1}

Previous infrared photometry and spectroscopy of UKS~1 giants revealed that it
is a very distant and reddened cluster, with distance modulus $\approx 15$, and
E(B-V) $\approx$ 3 \citep{Minniti95}, and moderately metal-rich, with [Fe/H]=-0.78
\citep{Origlia05}.

In order to determine the distance to UKS~1 we will use the horizontal branch
red clump, which is very conspicuous in this cluster. The red clump stars with
known parallaxes measured by the Hipparcos satellite are well calibrated
standard candles \citep[][]{Paczynski1998,Alves2000}. \cite{Alves2000}
obtained a K-band calibration of the horizontal branch red clump luminosity,
and applied the calibration to the red clump of the Galactic bulge. For
example, this calibration has also been applied to the red clump of the Large
Magellanic Cloud \citep[e.g.,][]{Alves2002,Pietrzynski,Borissova}.  The
uncertainties in the reddening should in all cases be larger than the
uncertainties due to the unknown metallicity.  Our $K_{\rm S}$ magnitudes are
in the 2MASS magnitude system. Therefore we transform the $K$-band
magnitudes of red clump stars of \cite{Alves2000} to $K_{\rm S}$ magnitudes
using $K=K_{\rm S} +0.044$ \citep[from][]{Grocholski}. The zero point
differences should be less than $0.02$ magnitudes \citep{Alves2002}. From the
previously cited works, we adopt the following mean values for the red clump
stars: $M_{K_s} = -1.65\pm0.03$, and $(J-K_s)_{0} = 0.71\pm0.03$
\citep[][]{lopez02}.

The distance modulus to the horizontal branch red clump in the globular cluster would be:

\begin{equation}
 \mu =  m_{K_s}- A_{K_s}/(A_J-A_{K_s}) [(J-K_s)-(J-K_s)_{0}] - M_{K_s}.
\end{equation}

Adopting the mean red clump magnitudes and colors discussed above, and the
reddening coefficients from \cite{Cardelli}, this simplifies to

\begin{equation}
 \mu = 5 log d(pc)- 5 = m_{K_s} - 0.73 (J-K_{s}) + 2.17.
\end{equation}

Using this equation we compute the mean distance modulus (and distance in kpc)
for the horizontal branch red clump in UKS~1, which has mean $K_{\rm S}=15.3$ and
$J-K_{\rm S}=2.0$ (Figure \ref{fig4}). This yields a mean $E(J-K)=1.30$, equivalent
to  $E(B-V)= 2.5$, and a distance modulus  $(m-M)_{0}= 16.01$,
  equivalent to 15.9 kpc. The uncertainty in the red clump position was
  estimated to be $\Delta$m$_{K_{\rm S}}$=0.012~$\mathrm {mag}$ and
  $\Delta(J-K_{\rm S})$=0.08~$\mathrm{mag}$, based on analysis of the relevant
  region of the CMD. This results in an error in the distance modulus
  and corresponding distance of $\Delta\mu$=0.078 and  $\Delta$D=0.6 kpc. The
distance to UKS~1 was also measured by \cite{Ortolani07} using deep CST
photometry. They find the mean magnitude of the horizontal branch at
$J=17.69$, and the mean colour $J-H=1.39 \mathrm {mag}$ or $1.71 \mathrm {mag}$ depending on the
calibration, and estimated $D=9.3$ to $14.3$ kpc, placing it beyond the
Galactic center. They find a total reddening of $E(B-V)=2.94 \mathrm {mag}$ to
$3.59 \mathrm {mag}$ also
depending on the calibration. Our values support the lower reddening values of
\cite{Ortolani07}. However, our distance measurement favors the largest
distance measurement of \cite{Ortolani07}, and it is consistent with the fact
that UKS~1 is located well beyond the Galactic center, but not as far as the
Sgr dwarf galaxy discovered by \cite{Ibata94}.

\section{The Globular Cluster Candidate VVV CL001}
\label{cl001}
Figure \ref{cmd} shows that the red giant branch of VVV CL001 is well defined,
but the precise location of the horizontal branch red clump is not, and we
cannot use the same method to estimate the distance. However, Figure
\ref{fig4} shows that the RGB of VVV~CL001 is bluer than that of UKS~1. This
may be due to a lower extinction or to a lower metallicity, or to a
combination of both effects. Within the selected sample, the saturation of the
brightest RGB stars prevents us from using the tip of the RGB for an
independent distance determination. The comparison between red clump and tip
of the RGB distance estimates could otherwise be used to assess the accuracy
of the distance. Although the absence of a populated horizontal branch red clump in
VVV~CL001 may be due to low metallicity, we test the
  possibility that this is a staticstical effect, due to the small sample. The CMDs of a randomly selected UKS~1 stars, corresponding
  to the VVV~CL001 sample size don't feature a distinct HB either. This leads us to the
  conclusion that the missing HB is mostly a statistical effect. Refing the
  selection criteria and an increased spatial resolution may help to detect
  some HB stars. 

Based on the CMD, and assuming similar reddenings for both clusters, we can
conclude that VVV~CL001 has similar metallicity or is slightly more metal-poor
than UKS~1. Otherwise it would be hard to reconcile the clearly differing
colours of the RGBs of these two clusters (see Figure \ref{cmd}), but they do
not allow for a large difference in metallicity.  The integrated
  near-infrared luminosity for Milky Way globular clusters are not very well
  known \citep[][and references therein]{cohen07}, nevertheless by comparing the RGB population in
  VVV~CL001 and UKS~1 we also estimate that the new cluster candidate is
  $\approx$3.7 mag fainter in $K_{\rm S}$ then UKS~1.

Thus, even though we do not have an accurate distance, the comparison with
UKS~1 allows us to conclude that VVV~CL001 is located beyond the Galactic
center, on the opposite side of the Milky Way. Therefore, this does not appear to be
a bulge globular cluster, as defined by \cite{Minniti95,Barbuy}.

For consistency, we can also determine the reddening from the mean location of
the VVV~CL001 red giants in the J-H vs $H-K_{\rm S}$ colour-colour diagram. The mean
$J-H=1.4$ and $H-K_{\rm S}$=0.4 yields $A_{\rm V}$=~6$\mathrm{mag}$ and $E(B-V)=~2.0$, 
slightly less reddened than UKS~1. On the other hand, 
the maps of \cite{Schlegel} give mean reddening of $E(B-V)\approx3.3$ for this
region of VVV field b351, which is larger than our measurements, but the reddening
towards the bulge is known to be nonuniform and patchy.

The data are not good enough to measure the distance, but we have an accurate
reddening and location, and deeper follow-up observations would allow to
define the other structural parameters of VVV~CL001. Based on the number of
giants we crudely estimate that this cluster has low mass, being about 50-100 times
less massive than UKS~1. In that case, VVV~CL001 would join the list of very
low luminosity globular clusters of the Milky Way \citep{koposov07}  (Table 1).
  Given the uncertainty of the cluster distance the possibility of VVV CL001
  being an old, compact open cluster is at this point not strictly ruled out,
  although the high stellar concentration and the well populated RGB favour
  the globular cluster interpretation. 

The proximity of these two globular clusters in the sky is surprising. There
are no other globular cluster pairs in the Milky Way separated by only $8$ arcmin.
If they are located at the same distance, and have the same radial velocity, these two clusters may be
bound: taking the distance of UKS~1 of $15.9$ $\pm$0.6~kpc, the separation would
be less than $40$ pc. There are physical pairs of clusters known in other
galaxies \citep[e.g.,][]{Bhatia,Dieball,Bekki,Minniti04}. However, the present
initial results are not good enough to indicate that VVV~CL001 is at the same
distance as UKS~1, and therefore we cannot claim that they form a physical binary cluster.
 We still need to measure their radial velocities to test the binary
  nature of the two clusters.

Alternatively, this may be just a chance alignment, and VVV~CL001 could be a
 globular cluster  or, although less likely, an old open cluster (see Section
 \ref{cl001}), either
more distant or closer to the Sun than UKS~1. Clearly, deeper photometric
observations are also needed for this object. 

\section{Conclusions}

\begin{figure*}[!ht]
\centering
\includegraphics[width=7.5cm]{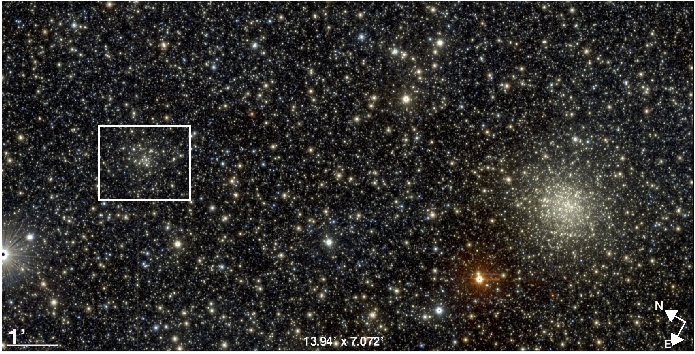}
\includegraphics[width=4.7cm]{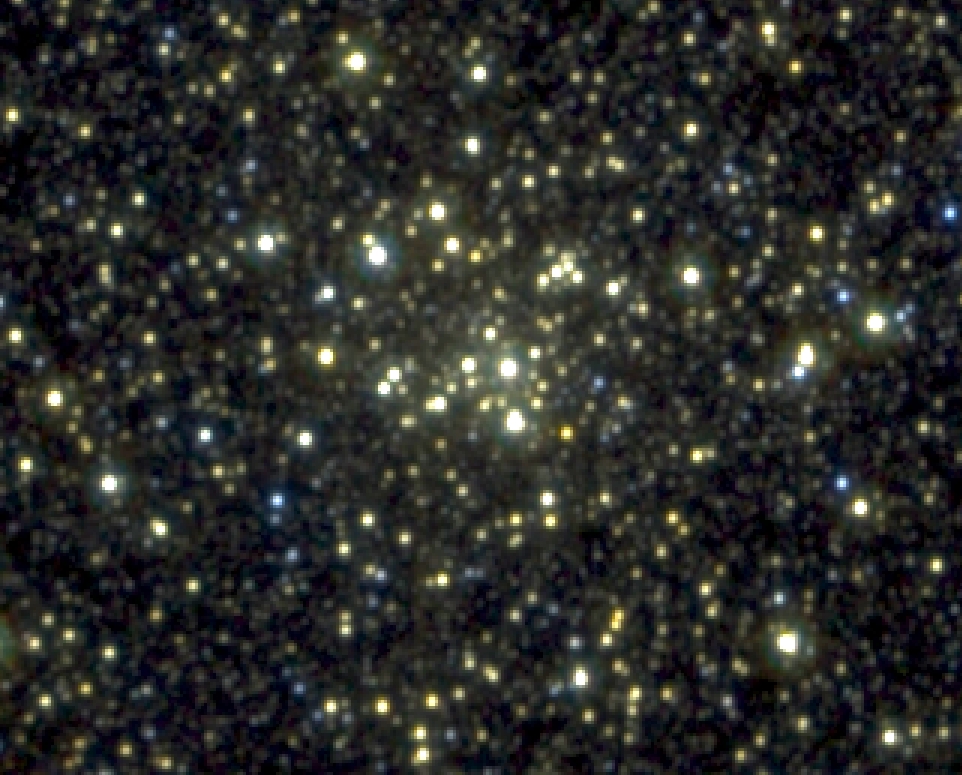}
  \caption{Discovery image (left panel, $JHK_{\rm S}$ colour composite) showing VVV~CL001
    on the left and UKS~1 on the right. The cluster UKS~1 was discovered by
    \cite{Malkan80}, with a total magnitude $I<16$ mag, and it was invisible
    in Palomar Survey plates, (and up to a few years ago UKS~1 was known as
    "the faintest globular cluster of the Milky Way"). The right panel
      shows a zoom into the region centered VVV~CL001, corresponding to the white box (1\farcm75 x
      1\farcm5).}
   \label{fig1}
\end{figure*}

\begin{figure}[!ht]
\includegraphics[width=8.5cm]{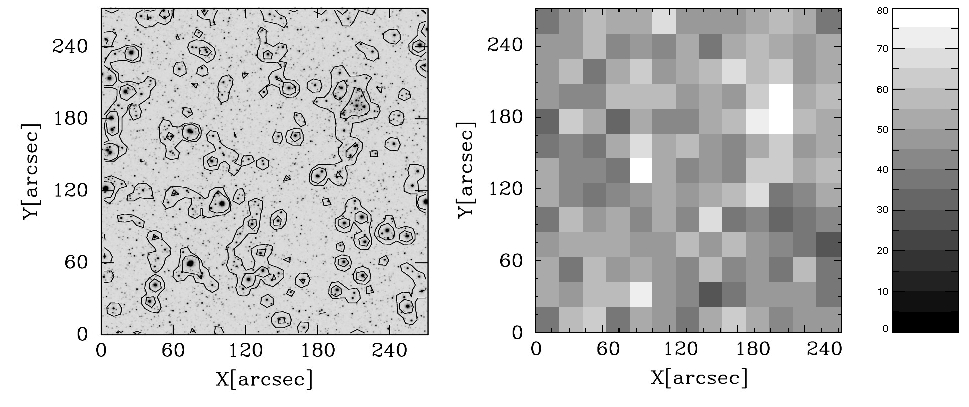}
\caption{J-band image of the VVV~CL001 region (left) and density of all
  objects. The contours in the left panel refer to the intensity level of the
  image. In contrast, to derive the significance of the over-density (right)
  only detections with a Dophot classification parameter 1 or 7
  \citep{Schechter} were selected. The sources were binned into 60x60 pixel
  bins (20 by 20 arcsec). We note that evenwith PSF photometry we are not able
  to resolve the inner region of VVV~CL001 (or UKS~1), which certainly affects
  the numerical results of our analysis (density, significance of the detected
  over-density). Both plots show the same 272 by 272 arcsec section of
    the J-band image. }
\label{density}
\end{figure}

The VVV Survey is searching for missing globular clusters in the inner regions
of the Milky Way galaxy. We report here the discovery of VVV CL001
(Figure \ref{fig1}), a low-mass globular cluster candidate at $\alpha=17:54:42.5$, 
$\delta=-24:00:53$. 
This is located only 8 arcmin away from the Galactic globular cluster
UKS~1 in the sky.

We take advantage of this spatial proximity to use UKS~1 as reference in order
to estimate the parameters of this newly discovered cluster. We measure the
distance and reddening of UKS~1, finding $E(B-V)_{UKS1}=2.5$ and $D_{UKS1}=15.9$kpc.

\begin{figure}[!ht]
  \includegraphics[height=4.5cm]{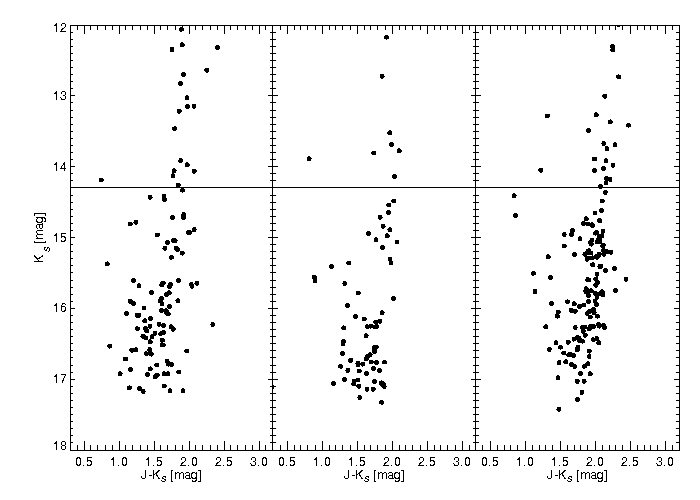}
  \caption{Discovery near-infrared colour-magnitude diagram of VVV~CL001 for a
    15 arcsec radius area centered on the cluster (left) compared with a
    field of same area located $3\farcm75$ east (middle), and with a
    field of UKS~1 of the same size and offset the cluster center by (45\farcs0) (right). The small offset for UKS~1 was applied,
    since the center of this cluster is barely resolved. The horizontal line at $Ks=14.3$ shows the
    limit of 2MASS photometry. Comparing UKS~1 and VVV~CL001
     we find that the RGB of UKS~1 is much more populated, and hence VVV~CL001
      to be far less massive then UKS~1.}
   \label{cmd}
\end{figure}

\begin{figure}[!ht]
  \includegraphics[height=4.5cm]{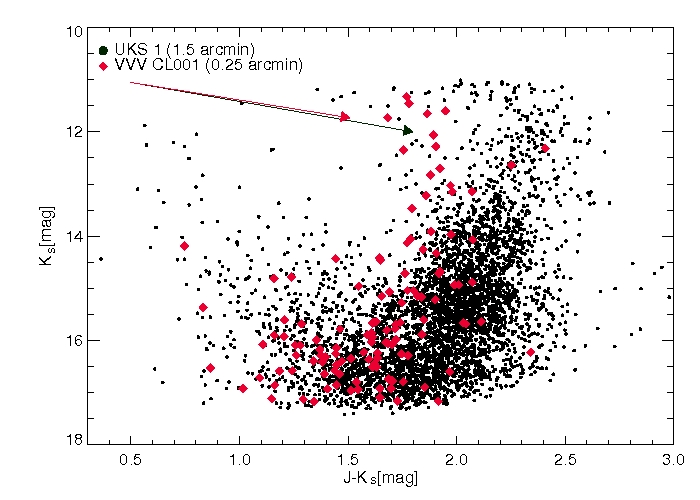}
  \caption{CMD of VVV~CL001 (diamonds) compared with UKS~1 (dots). The horizontal branch 
    red clump of UKS~1 is
    seen at $K_{\rm S}=15.3$, $J-K_{\rm S}=2.0$. This diagram shows that the RGB of VVV~CL001
    is slightly bluer than that of UKS~1, from which one can
    conclude that VVV~CL001 is less reddened or more metal-poor than UKS~1,
    and that the RGB of UKS~1 is much more populated, indicating the low mass 
    of VVV~CL001. The legend gives the radius of the selected
  cluster region, whereas the arrows indicate the reddening, derived for UKS~1.}
   \label{fig4}
\end{figure}

We present the first colour-magnitude diagrams of the new
cluster VVV~CL001 (Figure \ref{cmd}), estimating $E(B-V)= 2.0$.
We cannot define the mean $K_{\rm S}$-band magnitude of the horizontal branch red clump for this
cluster, and therefore its distance is uncertain. A very crude estimate by
comparison with the RGB of UKS~1 gives a similar distance, placing
VVV CL001 well beyond the bulge of the Milky Way. Observations in the
$K_{\rm S}$-band to be acquired in the following seasons by the VVV Survey would
allow us to improve this CMD of Figure \ref{cmd} and define the HB as
  well as the MSTO, which would allow a photometric age determination. Also, we estimate
that the RR Lyrae of VVV~CL001 would be within the limit of detection of our
VVV Survey.

We cannot definitely conclude that the proximity of UKS~1 and VVV CL001 on 
the sky imply that they compose a binary cluster because the distance to 
VVV CL001 is too uncertain. This remains as an exciting possibility that 
needs to be confirmed not only by means of a more accurate distance determination,
  but also by measuring their respective radial velocities from
  spectroscopic measurements. Finally, the present results are very encouraging and we conclude that the VVV
Survey can potentially provide the largest and most homogeneous census of globular
clusters in the survey area, out to well beyond the Galactic center, even in regions of large extinction.

\begin{table}
\caption[]{Properties of VVV~CL001, based on the VVV data, compared to two
  known extremely low luminosity globular clusters Koposov~1  \& ~2
  \citep[][]{koposov07}.\\
$^{*}$based on a distance of 16 kpc and a selected radius of 15 arcsec\\              }
\label{cluster}
\vskip 0.2cm
\begin{tabular}{l r r r}
\hline
\noalign{\smallskip}
   &   VVV CL001  &  Koposov~1 & Koposov~2 \\  

Position [l,b]   & 4.99$^\circ$, 1.84$^\circ$ & 260.98$^\circ$, 70.75$^\circ$ & 195.11$^\circ$, 25.55$^\circ$ \\
Distance [kpc]   & uncertain & $\approx$ 50kpc & $\approx$ 40 kpc\\
Radius $[arcmin]$ &  $\approx$1~pc $^{*}$& $\approx$~3~pc & $\approx$~3~pc\\

\noalign{\smallskip}
\hline
\end{tabular}
\end{table}

\begin{acknowledgements}
We thank the Cambridge Astronomical Survey Unit (CASU) for processing the
VISTA raw data. We acknowledge support by the FONDAP Center for Astrophysics 15010003, BASAL
Center for Astrophysics and Associated Technologies PFB-06, MILENIO Milky Way
Millennium Nucleus P07-021-F from MIDEPLAN, FONDECYT 1090213 from CONICYT, and
the European Southern Observatory. We use data products from the Two Micron
All Sky Survey, which is a joint project of the University of Massachusetts
and the Infrared Processing and Analysis Center/California Institute of
Technology, funded by the National Aeronautics and Space Administration and
the National Science Foundation. We also thank the referee, whose
  comments helped to improve the paper significantly.
\end{acknowledgements}

\bibliographystyle{aa} % style aa.bst
\addcontentsline{toc}{section}{\refname}\nocite{*}
\bibliography{Minniti} % your references Yourfile.bib

\end{document}